\documentclass[aps,groupedaddress,prd,twocolumn]{revtex4}
\usepackage{hyperref}

\usepackage{braket}

\usepackage{amsmath,amssymb,bm}
\usepackage{graphicx}
\usepackage{epstopdf}
\usepackage{amsfonts}
\usepackage{amssymb}
\usepackage{amsbsy}
\usepackage{amsmath}
\usepackage{array}
\usepackage{latexsym}
\usepackage{multirow}
\usepackage{sansmath}
\usepackage{subfigure}
\usepackage{lipsum}
\usepackage{float}
\usepackage{bm}
\usepackage{color}
\usepackage{csquotes}

\usepackage[plain]{algorithm}
\usepackage{algpseudocode}

\usepackage{xcolor}
\def\be{\begin{equation}}
\def\ee{\end{equation}}
\def\bea {\begin{eqnarray}}
\def\eea {\end{eqnarray}}

\makeatletter
\newcommand\makebig[2]{%
  \@xp\newcommand\@xp*\csname#1\endcsname{\bBigg@{#2}}%
  \@xp\newcommand\@xp*\csname#1l\endcsname{\@xp\mathopen\csname#1\endcsname}%
  \@xp\newcommand\@xp*\csname#1r\endcsname{\@xp\mathclose\csname#1\endcsname}%
}
\makeatother

\makebig{biggg} {3.0}
\makebig{Biggg} {3.5}
\makebig{bigggg}{4.0}
\makebig{Bigggg}{4.5}

\begin{document}

\title{Examining a Quantum System Subject to Energy Decoherence}

\author{Henry Crumley} \email{field.crumley@centre.edu}
\affiliation{Centre College,
Danville, KY, USA 40422}

\begin{abstract}
\vskip 0.2cm
This paper studies the energy decoherence of an interacting quantum system. It first reviews the experiments that motivated the postulates of quantum mechanics. It then discusses a decoherence that occurs dynamically in a closed system. This effect is studied in interacting quantum systems consisting of an oscillator and spins using analytical and numerical methods. The subsequent results are contrasted with cases with no decoherence. Connections of energy decoherence with thermodynamics are explored.
\end{abstract}

\maketitle
\setlength{\arrayrulewidth}{0.1mm}
\renewcommand{\arraystretch}{2}
\numberwithin{equation}{section}

This work studies the effects of energy decoherence on quantum degrees of freedom. For pedagogical completeness, I will introduce quantum theory, discuss how decoherence is conventionally accounted for in it, and then explain our work.

\section{Introduction}
Two experiments in particular inspired the development of quantum mechanics: the Ultraviolet Catastrophe and the double-slit experiment. 

According to classical theory, an ideal black body emits an infinite amount of energy at short wavelengths (in the ``ultraviolet'' range). This incongruity with nature is called the Ultraviolet Catastrophe. It was resolved by Max Planck, who proposed that radiation was emitted only at specific, quantized, wavelengths \cite{FC_MODPHY}. This argument pointed to a need for a new theory---one that allowed discrete, rather than continuous, energy levels to exist at small scales.

The second experiment that motivated quantum theory was the double-slit experiment. Consider two beams of particles, where each particle can only pass through one slit. Classical theory predicts that the produced intensity distribution by particles is binary---that is, two lines should appear, with the intensity pattern, $P_{1,2} = P_1 + P_2$. If instead we imagine a wave passing through the slits, then the two wavefunctions $\psi_1 $ and $\psi_2$ should emerge, producing the intensity pattern, $I_{1,2} = |\psi_1 + \psi_2|^2 $ \label{wave_combo}. When scientists sent one electron at a time through the slits, they expected to find a particle-like distribution. Instead, the final distribution was $\psi$-like. Scientists interpreted this behavior to mean that each electron possessed wavelike qualities \cite{FC_Shankar}.

\subsection{The Key Postulates of Quantum Mechanics}
Due to the aforementioned experiments, quantum mechanics was born. 

There are a few key differences between quantum and classical mechanics. First, in classical mechanics, one describes the state of a particle using the variables of position and momentum, that is, $x(t)$ and $p(t)$. These variables' values are always known, meaning that in an ideal experiment, both $x(t)$ and $p(t)$ could be determined simultaneously. We can also refer to $x(t)$ and $p(t)$ as phase space variables, whose evolution is described by Hamilton's equations \cite{taylorclassical},
\begin{align}
\dot{x} = \frac{\partial H}{\partial p} \quad \text{and} \quad \dot{p} = -\frac{\partial H}{\partial x}
\end{align}
where $H$ is the classical Hamiltonian. Now consider quantum theory: the state of a system is given by a normalized vector in a Hilbert space. The Hilbert space is spanned by basis elements. One possible basis are the eigenvectors of the Hamiltonian operator,
\begin{align}
\hat{H} \ket{n} = E_n \ket{n},
\end{align}
where $E_n$ are the allowed energy values. In this basis, a state can be expanded as
\begin{align}
    \ket\psi = \sum_nc_n\ket n,
\end{align}
where $c_n$ are complex values, and the Hamiltonian can be written as
\begin{align}
\hat{H} = \sum_n E_n \ket{n} \bra{n}.
\end{align}
Quantum evolution is given via the Schr\"{o}dinger equation,
\begin{align}
i\hbar\dot{\psi}=\hat{H}\psi.
\end{align}
For a time-independent Hamiltonian, the Schr\"{o}dinger equation has the solution
\begin{align}
\ket{\psi (t)} = \sum_n c_n e^{-i E_n t} \ket{\psi_n}.
\end{align}
To summarize, quantum mechanics describes a particle's state not by explicit position and momentum, but with a wavefunction $\ket{\psi}$ in a Hilbert space. Since we used the energy eigenbasis to span the Hilbert space, each dimension of the Hilbert space corresponds to an allowed energy level.

Classically, we can determine the exact value of $x$ and $p$. In quantum mechanics, we cannot. Instead, we are left with expectation values, that is, average values after many idealized measurements. We can determine these values by calculating the inner product (the complex-valued version of a dot product \cite{FC_griffiths, FC_Shankar}), where  $\hat{X}$ and $\hat{P}$ are linear operators (matrices). Letting $\hat O$ be a dummy operator, and considering a system in a state $\psi$, we can describe the average value of $\hat O$ as,
\begin{align}
\braket{\hat{O}} = \braket{\psi | \hat{O} | \psi}.
\end{align}
Note that this calculation first transforms the state, $\hat{O}\ket{\psi} = \ket{\psi_O}$, and then takes its inner product, $\braket{\psi|\psi_O} = \braket{O}$. Thus, a measurement changes the original vector's state from $\ket{\psi}$ to $\ket{\psi_O}$ \cite{FC_Shankar}. The uncertainty value of an operator is determined via the equation,
\begin{align}
    \sigma_{O} = \sqrt{\braket{\hat O^2}-\braket{\hat O}^2}.
    \label{uncertainty}
\end{align}
Different measurements may change the state differently. This is why we in general do not expect operations to commute. One example is position and momentum. Indeed it can be found that in general 
\begin{align}
[\hat{X}, \hat{P}] = i\hbar.
\end{align}
This ``Canonical Commutation Relation" leads to the ``Heisenberg Uncertainty Principle," that is,
\begin{align}
    \sigma_{x}\sigma_{p}\geq\frac{\hbar}{2}
    \label{heisenberg}
\end{align}
\cite{FC_Shankar}.

Finally, recall that in classical mechanics, one combines independent wavefunctions using addition:
\begin{align}
    \psi_{classical} = |\psi_1+\psi_2+\dots +\psi_n|
\end{align}
In quantum mechanics, this process is performed using the tensor product, an operation that joins multiple Hilbert spaces, enabling us to define joint states and joint operators. For instance consider two quantum systems, $\psi_A $ and $\psi_B $. Then,
\begin{align}
\text{if }\psi_A &= a\ket{0} + b\ket{1}\\
\text{and }\psi_B &= c\ket{0} + d\ket{1},\\
\text{the state } \psi_{AB} &= \psi_A \otimes \psi_B\\   
\text{is }\psi_{AB} &= ac\ket{00} + ad\ket{01} + bc\ket{10} + bd\ket{11}.
\end{align}
Thus, the tensor product stores every linear combination of the two states $\psi_A$ and $\psi_B$ in a combined state, $\psi_{AB}$. Joining two operators is a similar procedure. 

This formulation of quantum theory captures the expected discreteness at microscopic scales. Next, we will explore a formulation of quantum theory that will enable us to describe the aforementioned interference.

\subsection{Density Matrix Formalism}
So far, we have considered the vector formalism of quantum mechanics. We will hereafter use the density matrix formalism, $\hat\rho = \ket\psi\bra\psi$. There is a key reason for this change. Consider a state
\begin{align}
    \ket\psi = \sum_nc_n\ket n.
    \label{arb_state}
\end{align}
If we let our basis be two-dimensional, then $\hat \rho$  is
\begin{align}
    \hat\rho = \ket{\psi}\bra{\psi} = \sum_{m,n}c_mc_n^*\ket{m}\bra{n} = \begin{bmatrix}
        c_{00} & c_{01} \\
        c_{10} & c_{11}
    \end{bmatrix},
\end{align}
using the convention $c_i^*c_j = c_{ij}$. The probability of a state being in the allowed energy levels $c_0$ and $c_1$ is represented by $c_{00}$ and $c_{11}$ ($c_{nn}$ is equivalent to $|c_n|^2$). The cross-terms $c_{01}$ and $c_{10}$ represent probabilities of the system being found in a combination of two states. These are called interference states, states that are never measured, but whose influence causes the wavelike properties of a quantum system. Hence, $\hat\rho$ stores all possible interference states of $\ket\psi$ in a matrix. Since decoherence destroys interference states, and our project studies energy decoherence, using $\hat\rho$ rather than $\ket\psi$ enables us to directly observe how a loss of interference states affects a quantum system.
\label{decoherence_mixing}

Note that $\hat\rho$ is a matrix. Following from the definition of the density matrix,
\begin{itemize}
    \item $\hat\rho^\dagger = \hat\rho$ (the matrix is symmetric),
    \item Trace$(\hat\rho) = 1$ (since $\ket\psi$ is normalized),
    \item and Trace$(\hat\rho^2) \leq 1$ (the relation is an equality when $\hat\rho$ is pure).
\end{itemize}
Note the mention of a pure state in the above bullet point. A pure state can be rewritten as the vector matrix product $\hat\rho(t) = \ket{\psi_t}\bra{\psi_t}$ after time evolving in $\hat\rho$ form. If this cannot be done, then we say that the state is mixed \cite{FC_Shankar}. Since one form of mixing is the vanishing of the off diagonal terms, destruction of interference---i.e. decoherence---causes mixed states. Note that since $\hat\rho^\dagger = \hat\rho$, the matrix is symmetric. This will remain true even with a mixed state. Thus, we will always retain the statistical information of $\hat\rho$, even when it is incongruous with the vector form.

All of the standard processes in quantum theory can be rephrased with this formalism. The time derivative of the state $\hat{\dot\rho}$ is now described by the von Neumann equation:
\begin{align}
    \hat{\dot\rho}= -i[\hat H,\hat\rho],
\end{align}
where $\hat H$ is unchanged from the $\psi$ formalism. The solution to the von Neumann equation is
\begin{align}
    \hat\rho (t) = \sum_{m,n}c_mc_n^*e^{-i t (E_m-E_n)}\ket m\bra n,
\end{align}
Thus, we can see that standard quantum evolution does not destroy the interference patterns. To determine the expectation value of an operator, we now take the trace of the operator and the state at that time:
\begin{align}
    \braket{\hat O} = \text{Trace}(\hat O\hat\rho).
\end{align}

\subsection{Motivation for the Project}
Decoherence normally emerges by introducing ``environmental" degrees of freedom to the system. In these scenarios, the decoherence occurs after enough collisions on a system (each collision transforms the state) occur, so that, prior to measurement, the interference states have been sufficiently eliminated \cite{FC_Decoh_justification}. For our system, we will study decoherence that occurs dynamically in a closed system, without introducing the aforementioned environmental degrees of freedom (for more details, see \cite{FC_Time_uncert_spot}).

\section{Modified evolution}
Our energy decoherence occurs dynamically. A common model for this is via a modified von Neuman equation: 
\begin{align} 
    \hat{\dot\rho} = -i[\hat H,\hat\rho]-\frac{\tau}{2}[\hat H,[\hat H,\hat\rho]],
\end{align}
where $\tau$ indicates the the strength of decoherence \cite{FC_Time_uncert_spot}. The solution to the modified von Neumann equation is
\begin{align}
    \hat\rho (t) = \sum_{m,n}c_mc_n^*e^{-i(E_m-E_n)t}e^{-\frac{\tau(E_m-E_n)^2t}{2}}\ket m\bra n .
    \label{solution}
\end{align}
The vanishing of the off-diagonal terms at late times indicates energy decoherence. Consider what happens at large $t$. Setting $t\rightarrow\infty$ sends all of $\hat\rho$ to zero, except for the terms where $E_n = E_m$, that is, the diagonal terms, where the exponential disappears, and we are left with 
\begin{align}
    \hat\rho(t\rightarrow\infty) = \sum_n|c_n|^2\ket{n}\bra{n}.
\end{align}
Thus, at large times, only the diagonals of $\hat\rho$ are nonzero, and we have a decohered state. Note that this decoherence will happen in all states, regardless of the Hamiltonian $\hat H$, or the magnitude of $\tau$.

Next, we considered the trace of $\hat\rho^2$ at large $t$. Again, the exponential disappears, and we are left with the diagonal matrix values,
\begin{align}
    \text{Trace} (\hat{\rho}^2(t\rightarrow\infty)) = \sum_{m}|c_{m}|^4.
\end{align}
Note that a maximally mixed state $\hat\rho$ is one where this sum is closest to zero. Thus, to maximally mix a state $\hat\rho$, we need equal $c_m$ values. Entropy can be defined as
\begin{align}
    S = \text{Trace}(\hat\rho\log\hat\rho) = 1-\text{Trace}(\hat\rho^2).
    \label{entropy}
\end{align}
Thus, the maximal entropy of a quantum $N$-dimensional system is one with entries $c_n = \frac{1}{\sqrt{N}}$. Our mixing occurs dynamically. Thus, the entropy of the states undergoing decoherence is expected to increase over time. This bears resemblance to the 2nd law of thermodynamics, a point that will be revisited later.

To determine the relation between the level of information at $\ket{\psi_0}$ and the final entropy of the system after evolution, we considered random initial values for the state, evolved the system until $\tau$ dominated, and performed the distance measure:
\begin{align}
    D = \sqrt{\sum_{i = 1}^N||c_i|-\frac{1}{\sqrt{N}}|^2}
\end{align}
Comparing this distance measure to the state's final entropy after decoherence clearly demonstrates the two's relationship.

\begin{figure}[h]
    \centering
    \includegraphics[width=0.4\textwidth]{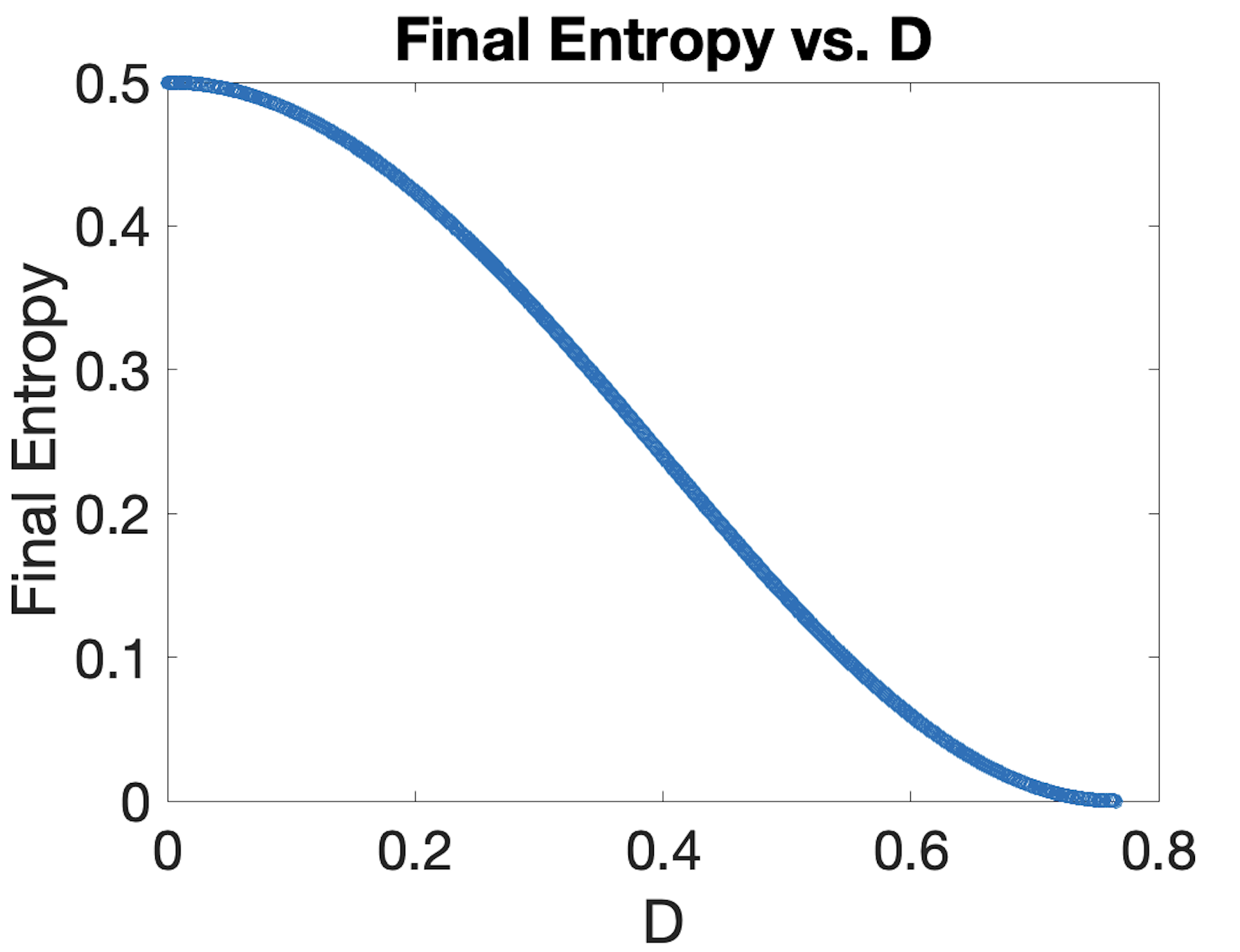} 
    \caption{Difference from initial optimal mixing level compared with final entropy for the simple harmonic oscillator. $10^5$ random initial states of the form $\ket{\psi_0} = c_0\ket{0}+c_1\ket{1}$ in a $2$ dimensional Hilbert Space were used to generate the plot.}
\end{figure}
\section{Application}
\subsection{Single quantum system}
The single system we consider is the quantum oscillator. Since we are studying energy decoherence, a natural basis to use is the energy basis $\ket{n}$, which has fixed energy levels $E_n=(n+1/2)\hbar\omega$. 
We use the creation and annihilation operators to transition between energy levels. These are:
\begin{align}
   \hat a\ket{n} = \sqrt{n}\ket{n-1}, \\
    \hat a^\dagger\ket{n} = \sqrt{n+1}\ket{n+1}.
\end{align}
In terms of the creation and annihilation operators, the Hamiltonian is \cite{FC_Shankar}
\begin{align}
    \hat H = \hat a^\dagger \hat a +\frac{1}{2}I,
\end{align}
and the position and momentum operators are \cite{FC_Shankar}:
\begin{align}
    \hat x = \sqrt{\frac{\hbar}{2m\omega}}(\hat a+\hat a^\dagger);\quad
    \hat p = i\sqrt{\frac{
    \hbar m\omega}{2}}(\hat a^\dagger - \hat a).
\end{align}
Based upon our previous solution (see \ref{solution}) we find that
 \begin{widetext}
 \begin{equation}
    \braket{\hat x} = \sqrt{\frac{\hbar}{2m\omega}} e^{-\frac{\tau(\hbar\omega)^2}{2}t} \Big( e^{-i\hbar\omega t} \sum_l c_l^* c_{l+1} \sqrt{l+1} + e^{i\hbar\omega t} \sum_l c_l^* c_{l-1} \sqrt{l} \Big),
 \end{equation}
 \\
 \begin{equation}
    \braket{\hat p}= i \sqrt{\frac{\hbar m\omega}{2}} e^{-\frac{\tau(\hbar\omega)^2}{2}t} \Big( e^{-i\hbar\omega t} \sum_l c_l^* c_{l+1} \sqrt{l+1} - e^{i\hbar\omega t} \sum_l c_l^* c_{l-1} \sqrt{l} \Big),
 \end{equation}
 \\
 \label{soln}
 \begin{equation}
\braket{\hat x^2} = \frac{\hbar}{2m\omega}\sum_n c_nc_{n-2}^*e^{2i\hbar\omega t - 2\tau\hbar^2\omega^2t}\sqrt{n^2-n}+c_nc_{n+2}^*e^{-2i\hbar\omega t - 2\tau\hbar^2\omega^2t}\sqrt{(n+1)(n+2)}+ n|c_n|^2+ (n+1)|c_n|^2,
 \end{equation}
 \\
 \begin{equation}
\braket{\hat p^2} = -\frac{\hbar m\omega}{2}\sum_n c_nc_{n-2}^*e^{2i\hbar\omega t - 2\tau\hbar^2\omega^2t}\sqrt{n^2-n}+c_nc_{n+2}^*e^{-2i\hbar\omega t - 2\tau\hbar^2\omega^2t}\sqrt{(n+1)(n+2)} - n|c_n|^2 - (n+1)|c_n|^2.
 \end{equation}
\end{widetext}
Using our analytical solutions, we can calculate the uncertainty values for position and momentum (see \ref{uncertainty}) once $t$ is large and $\tau$ dominates. Taking the limit $t\rightarrow\infty$,
\begin{align}
     \lim_{t\rightarrow\infty}\braket{\hat x} = 0; \quad \lim_{t\rightarrow\infty}\braket{\hat x^2} = \frac{\hbar}{2m\omega}\sum_n (2n+1)|c_n|^2,
\end{align}
and
\begin{align}
    \lim_{t\rightarrow\infty}\braket{\hat p} = 0; \quad \lim_{t\rightarrow\infty}\braket{\hat p^2} = \frac{\hbar m\omega}{2}\sum_n (2n+1)|c_n|^2.
    \label{uncert_justification}
\end{align}
Before proceeding, note that as $t\rightarrow\infty$, $\braket{\hat x} = \braket{\hat p} = 0$ for all $\tau$. This is due to a time dependence of all terms in the solution. Note too that in the limit $\braket{\hat x^2}$ and $\braket{\hat p^2}$ are nonzero. Thus, there is a nonzero probability of measuring the oscillator in a nonzero position and momentum when $t$ is large.

Next, we calculate $\sigma_{x}$ and $\sigma_{p}$:
\begin{align}
    \sigma_{x} = \sqrt{\frac{\hbar}{2m\omega}\sum_n (2n+1)|c_n|^2} \\
    \sigma_{p} = \sqrt{\frac{\hbar m\omega}{2}\sum_n (2n+1)|c_n|^2}.
\end{align}
So,
\begin{align}
    \sigma_{x}\sigma_{p} \geq \frac{\hbar}{2}.
\end{align}
The Heisenberg Uncertainty Principle (see \ref {uncertainty}, \ref{heisenberg}) is preserved with the right hand side equaling $\frac{\hbar}{2}$ when $\ket\psi$ is fully in the ground state.

The results above are applicable to a general state (any state of the form \ref{arb_state}). We will now specialize to the coherent state. The coherent state in quantum mechanics is designed to imitate classical behavior in that it minimizes the amount of uncertainty in the system (the Heisenberg Uncertainty Principle (see \ref{heisenberg}) is fixed as an equality, rather than a greater-or-equal-to statement)\cite{Hall_ND}. Hence, an oscillator in this state is most similar to a classical oscillator. For our oscillator to be in the coherent state, its initial position and momentum must adhere to:
\begin{align}
    \ket{\phi_0} = \cos\left(\frac{\theta}{2}\right)\ket{0} + \sin\left(\frac{\theta}{2}\right)e^{i\phi}\ket{1},
\end{align}
where 
\begin{align}
    x_0 = \braket{\hat x}(t = 0) = \frac{\sin\theta\cos\phi}{\sqrt{2m\omega}},
\end{align}
and
\begin{align}
    p_0= \braket{\hat p}(t = 0) = -\sqrt{\frac{m\omega}{2}}\sin\phi\sin\theta,
\end{align}
for allowed $\theta$ and $\phi$ values, as done in \cite{FC_Hamiltonian_Ideas}. Thus, we set initial values to $\omega = \sqrt{\frac{k}{m}} = 1/2$, $p_0 = 0$, and $x_0 = \frac{1}{3}$. 

\begin{figure}[h]
    \centering
    \includegraphics[width=0.47\textwidth]{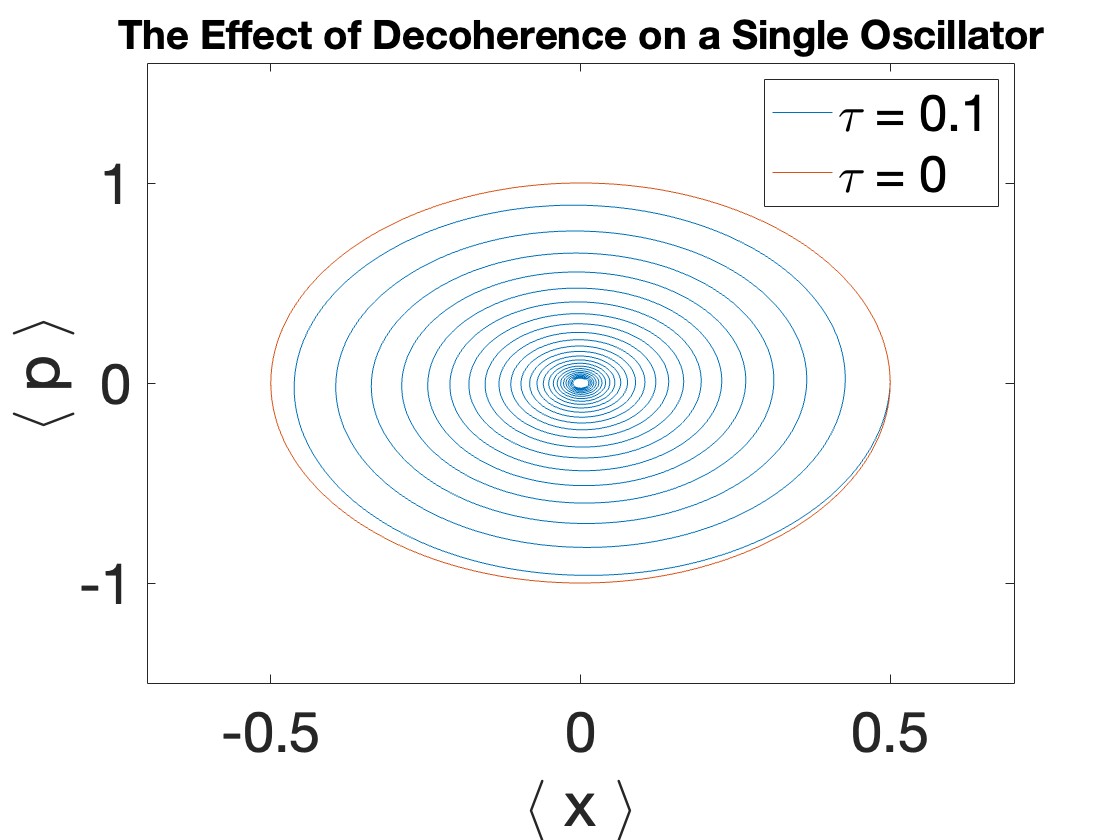} 
    \caption{Phase plot for the expectation values of a single oscillator with and without decoherence. $\tau = 0.1, k = 1, \omega = \frac{1}{2}, T = 300$. Although $\braket{\hat x}$ and $\braket{\hat p}$ go to zero at large $t$, the Heisenberg Uncertainty Principle is still followed, meaning there is a probability of measuring $\ket\psi$ and finding it in a non-$\braket{\hat x},\braket{\hat p} = 0$ state at large $t$ (see \ref{uncert_justification}).}
    \label{phase_coherent}
\end{figure}

We can reflect upon the oscillator's behavior in a phase plot. Note that, as predicted, at large timescales, the decoherent model's expectation values for position and momentum both approach zero (see figure \ref{phase_coherent}).
We also solved for $\braket{\hat H}$ and entropy. For $\braket{\hat H}$, we found
\begin{align}
    \braket{\hat H} =\hbar\omega\sum_m |c_m|^2 (m+1/2).
    \label{conservation}
\end{align}
Thus, for all Hamiltonians, energy in the system will be conserved. Based upon \ref{entropy}, entropy was determined to be:
\begin{align}
    S(t) = 1 - \sum_{m,n}|c_m|^2|c_n|^2e^{-\tau(E_n-E_m)^2t},
\end{align}
showing an energy and time dependence. 

\begin{figure}[h]
    \centering
    \includegraphics[width=0.47\textwidth]{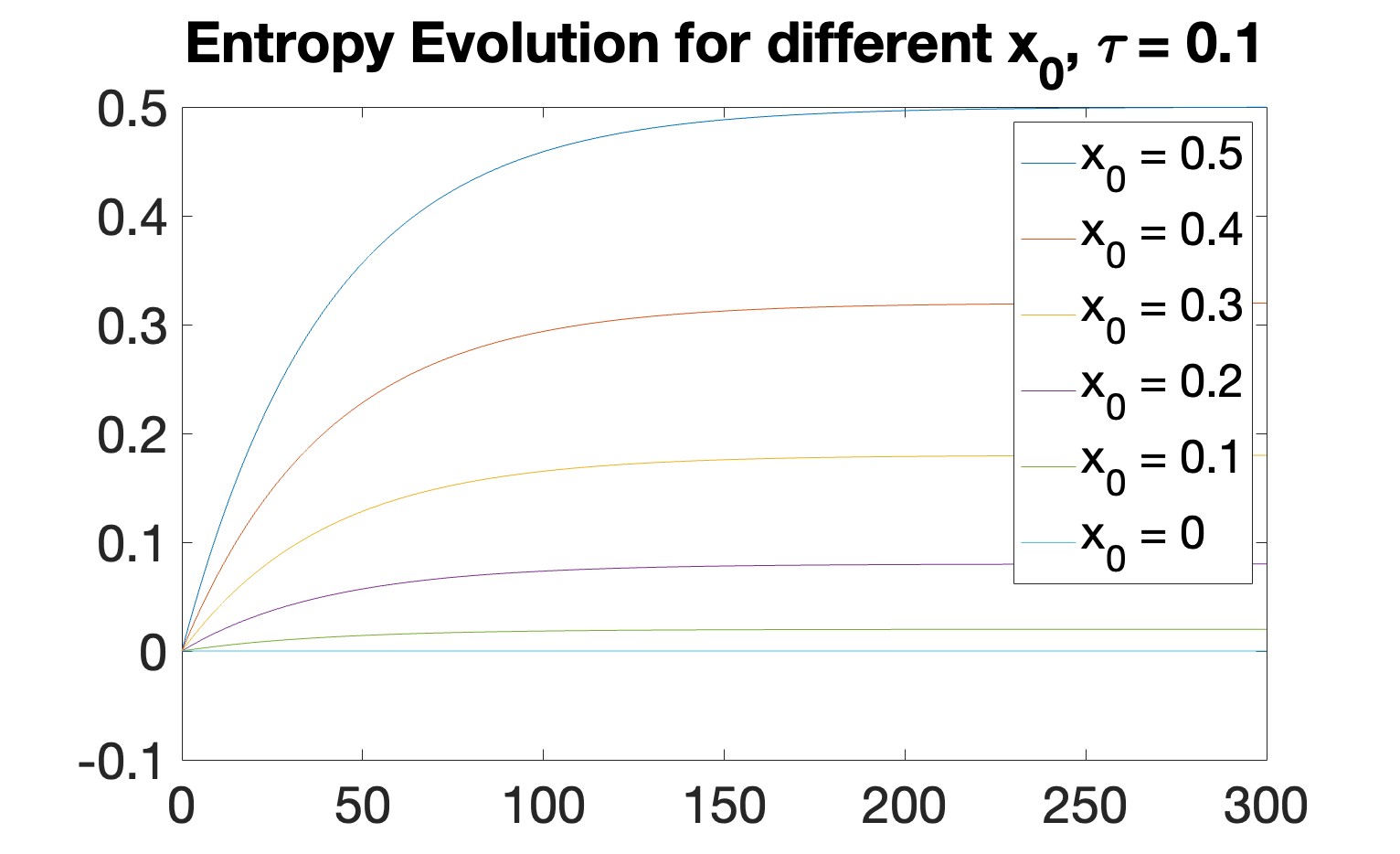} 
    \caption{Time evolution of entropy for different initial oscillator positions. $\tau = 0.1, k = 1, \omega = \frac{1}{2}, T = 300$.}
    \label{diff_init_pos}
\end{figure}

We also examined the relationship between different initial oscillator positions and the final entropy of the system. Initial positions set at a greater distance from the center of oscillation led to larger final entropy values. This can be understood as a result of a higher amount of energy being present in these systems (see figure \ref{diff_init_pos}), and bears a resemblance to the first law of thermodynamics, a point we will revisit in a later section.

\subsection{Interacting quantum system}
We next investigated two more complex quantum systems---a tripartite (oscillator-spin-spin) system, and coupled oscillators.
\begin{figure}[h]
    \centering
    \includegraphics[width=0.48\textwidth]{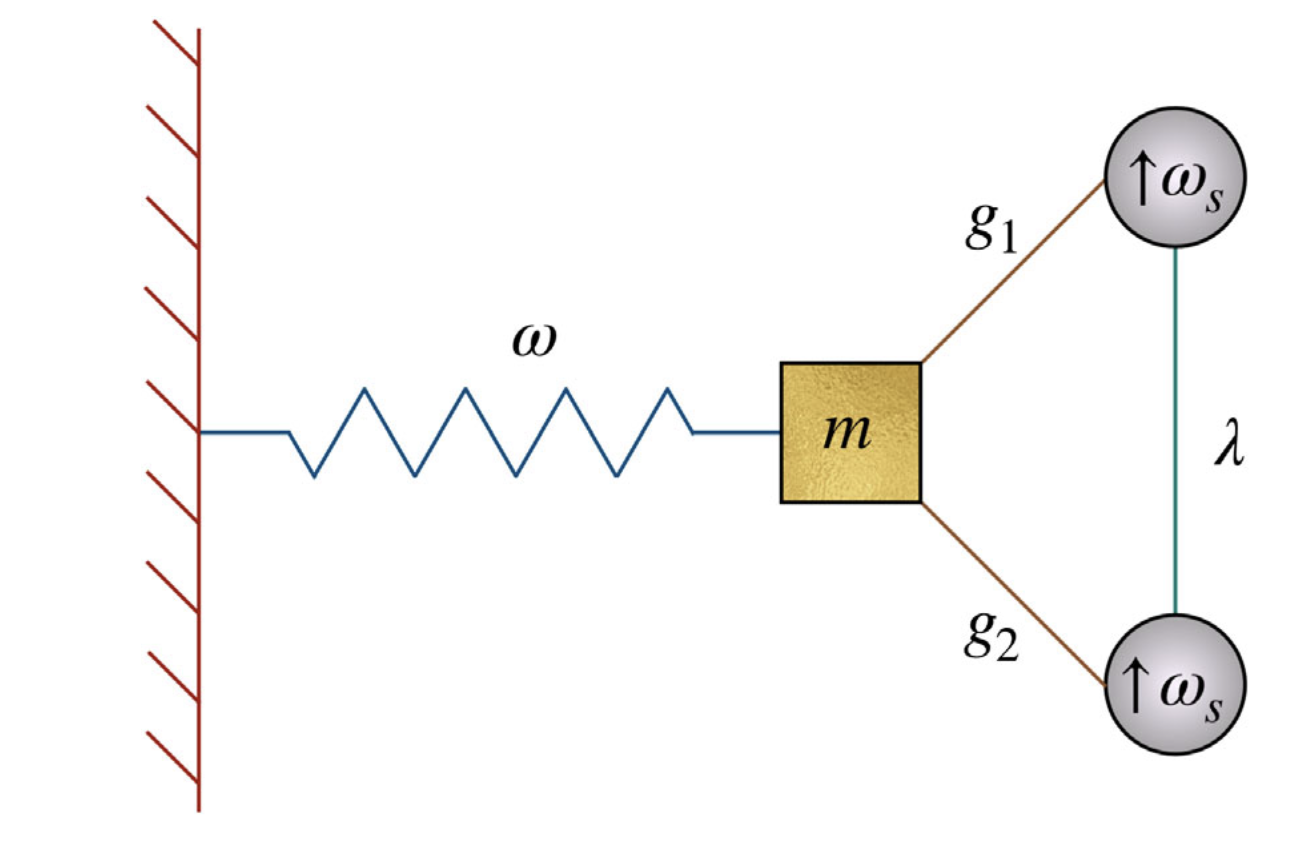} 
    \caption{Diagram of the tripartite system. Spins act like an energy reservoir for the oscillator. For all figures, the spin states were set to up, up. coupling and $w_s$ values were either $1$ or $0$. Figure taken from \cite{FC_Hamiltonian_Ideas}.}
    \label{spins}
\end{figure}
In our tripartite system, we coupled an oscillator to two coupled spins (as shown in figure \ref{spins}). Each spin can either be measured in a state  $\ket{\uparrow} = \ket{1} = \begin{bmatrix} 1 \\ 0 \end{bmatrix}$ (spin up) or $\ket{\downarrow} = \ket{0} = \begin{bmatrix} 0 \\ 1 \end{bmatrix}$ (spin down). Thus, a single spin's Hilbert Space is two-dimensional. Its operators are the Pauli Spin Matrices \cite{FC_Shankar}, 
\begin{align}
    \hat \sigma_x = \begin{bmatrix} 0 & 1 \\ 1 & 0 \end{bmatrix};\quad \hat \sigma_y = \begin{bmatrix} 0 & -i \\ i & 0 \end{bmatrix};\quad \hat \sigma_z = \begin{bmatrix} 1 & 0 \\ 0 & -1 \end{bmatrix}. 
\end{align}
Note that the matrices $\hat \sigma_{x,y,z}$ have no relation to the terms $\sigma_x$ and $\sigma_p$ that pertained to uncertainty in \ref{uncertainty}. Similar to the raising and lowering operators $\hat a$ and $\hat a^\dagger$, the spins' energy levels can be raised and lowered using the operators
\begin{align}
\hat \sigma_+ = \frac{1}{2} (\hat \sigma_x + i \hat \sigma_y); \\
\hat \sigma_- = \frac{1}{2} (\hat \sigma_x - i \hat \sigma_y),
\end{align}
where $\hat \sigma_+$ changes $\ket\downarrow$ to $\ket\uparrow$, and maintains $\ket{\uparrow}$ as $\ket\uparrow$, and $\hat \sigma_-$ does the opposite \cite{FC_griffiths}.

We will now proceed by describing the Hamiltonian $\hat H$ of the system. This Hamiltonian has four components, $\hat H = \hat h_o+\hat h_s+\hat h_{g1}+\hat h_{g2}+\hat h_{\lambda}$, that is, the addition of the isolated oscillator, isolated spins, coupled oscillator and spin, and coupled spin components \cite{FC_Hamiltonian_Ideas}:

\begin{align}
\hat h_o = (\frac{p^2}{2m}+\frac{1}{2}m\omega^2\hat x^2)\otimes(I^{(1)}\otimes I^{(2)})
\end{align}

\begin{align}
\hat h_s  = I\otimes\frac{\omega_s}{2}(\hat\sigma_z^{(1)}\otimes I^{(2)}+I^{(1)}\otimes\hat\sigma_z^{(2)})
\end{align}

\begin{align}
\hat h_{g1} = \frac{g_1}{2}(\hat a\otimes\hat\sigma_+^{(1)}+\hat a^\dagger\otimes\hat\sigma_-^{(1)})\otimes I^{(2)}
\end{align}

\begin{align}
\hat h_{g2} = \frac{g_2}{2}(\hat a\otimes I^{(1)}\otimes\hat\sigma_+^{(2)}+\hat a^\dagger\otimes I^{(1)}\otimes\hat\sigma_-^{(2)})
\end{align}
and
\begin{align}
\hat h_{\lambda} = \frac{\lambda}{2}I\otimes(\hat\sigma_+^{(1)}\otimes\hat\sigma_-^{(2)}+\hat\sigma_-^{(1)}\otimes\hat\sigma_+^{(2)}).
\end{align}

The identity elements are present wherever there is no term dependence. For instance, $\hat h_{\lambda}$ characterizes spin-spin coupling, but does nothing to the oscillator, hence it acts as an identity on the oscillator's Hilbert space.

\begin{figure}[h]
    \centering
    \includegraphics[width=0.48\textwidth]{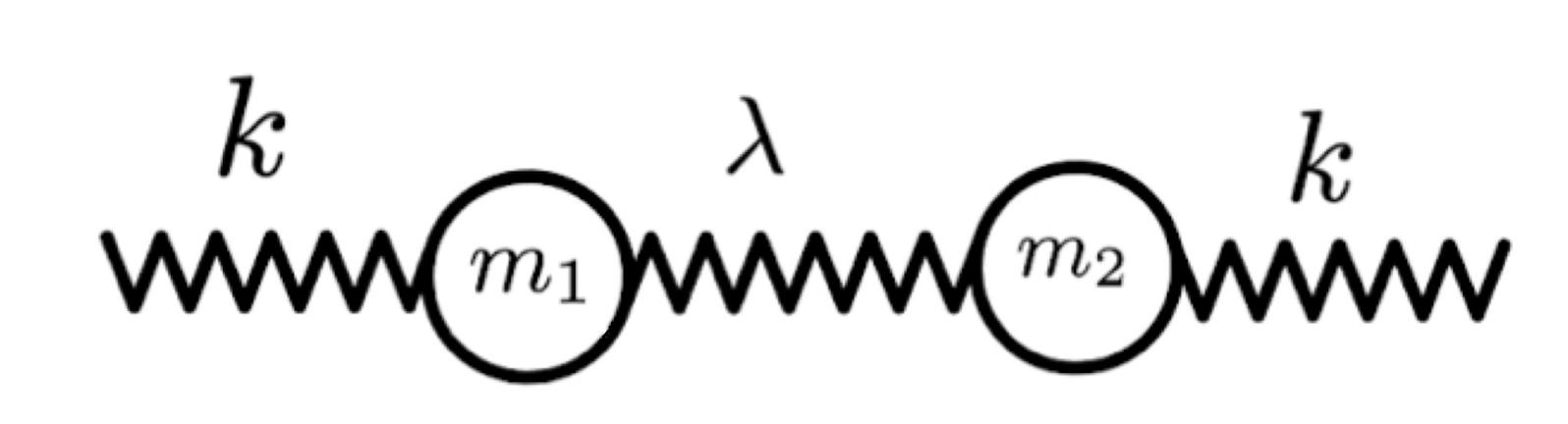} 
    \caption{Diagram of the coupled oscillators. The spring constants $k$ were set to $1$, $m_1 = 4$, $m_2 = 1$, and $\lambda = 1$.}
    \label{osc}
\end{figure}

\begin{figure}[h]
    \centering
    \includegraphics[width=0.23\textwidth]{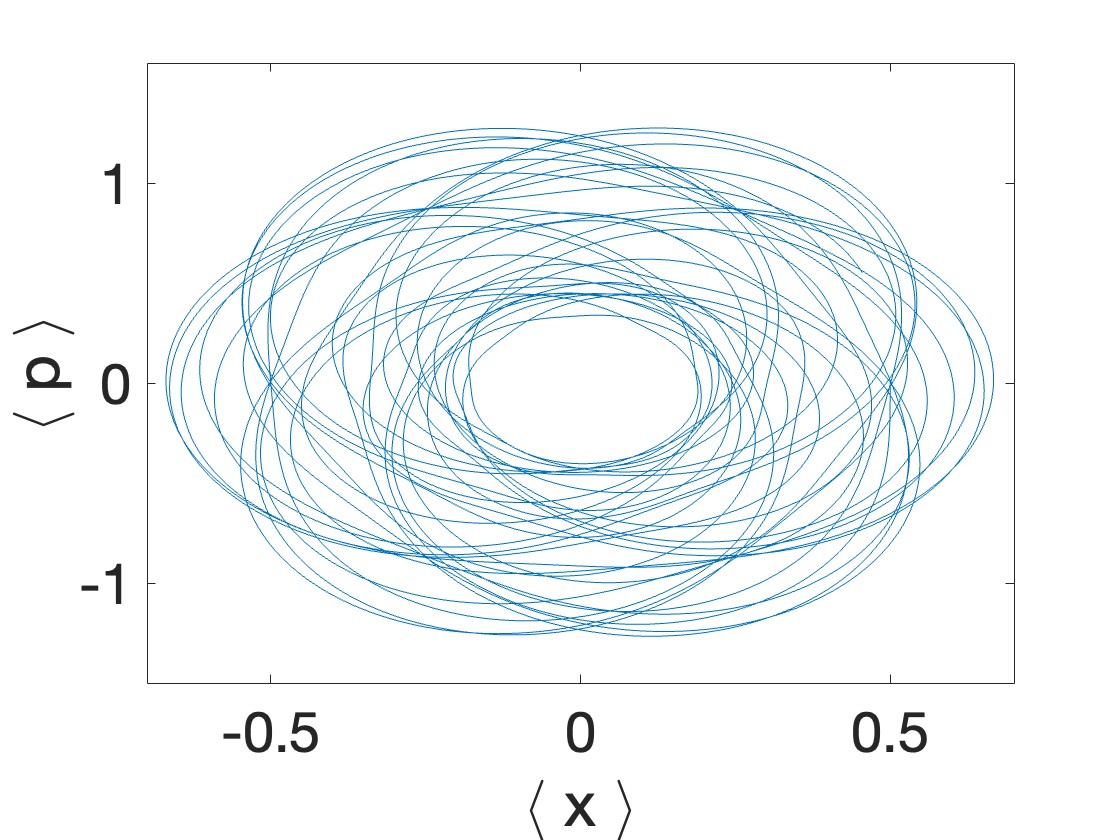} 
    \includegraphics[width=0.23\textwidth]{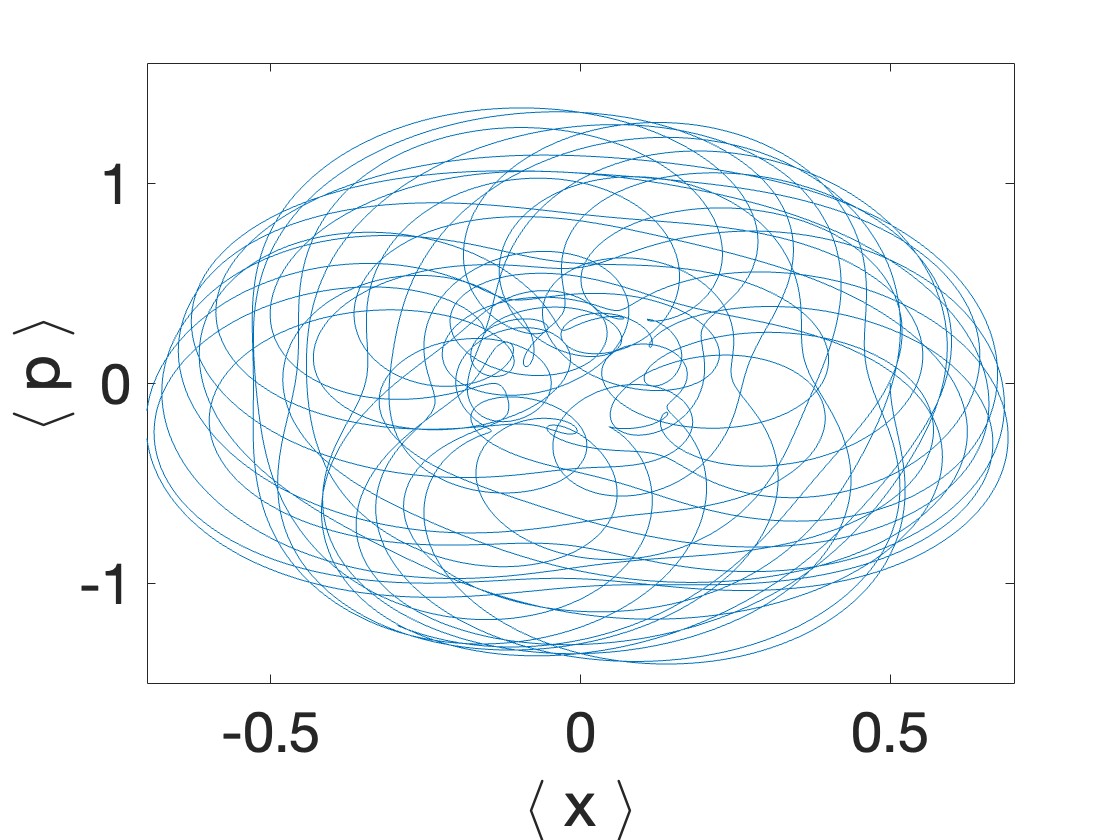} 
    \includegraphics[width=0.23\textwidth]{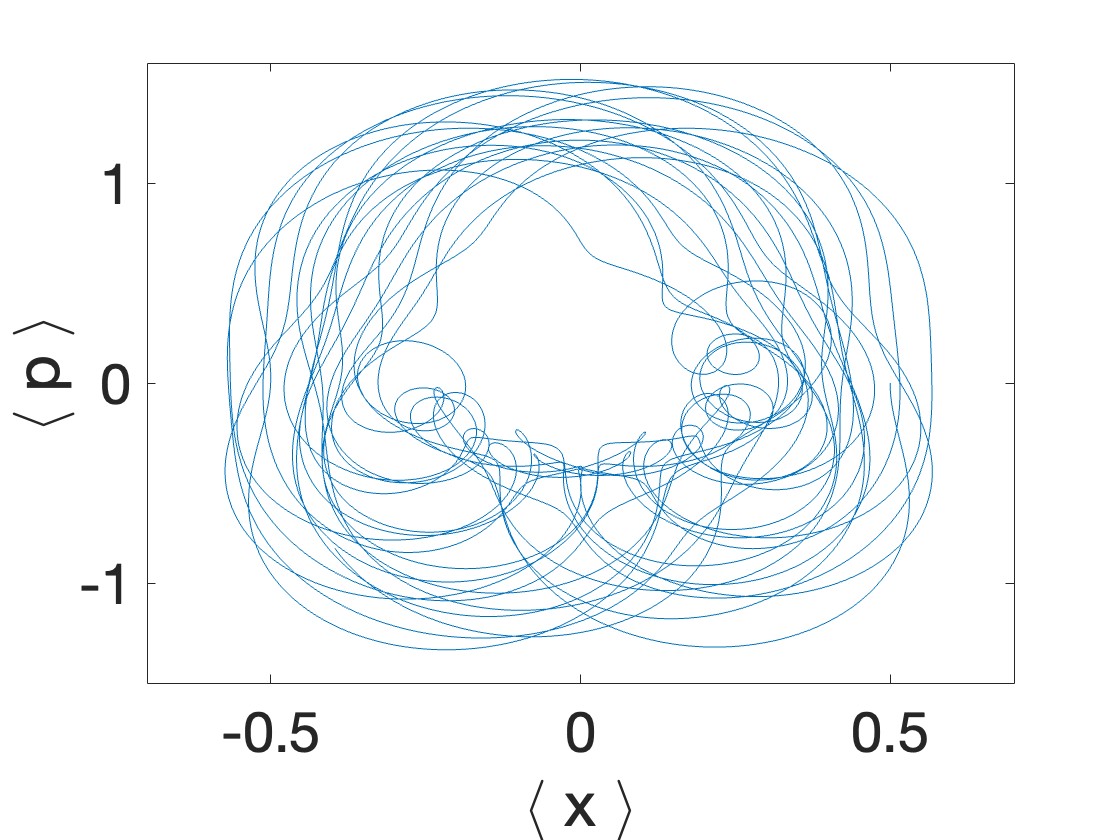} 
    \includegraphics[width=0.23\textwidth]{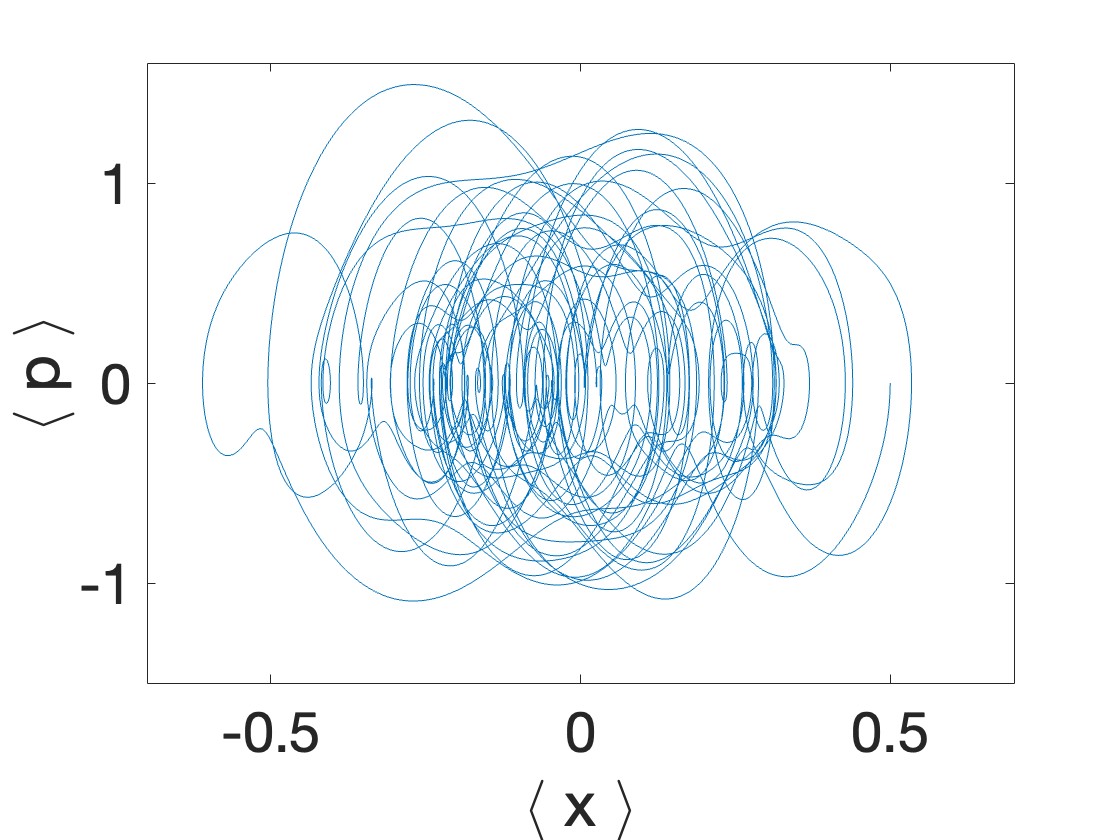} 
    \caption{Phase plot of a traced-out oscillator for the more complicated systems, without decoherence ($\tau = 0, x_0 = 0.5, p_0 = 0$). The spring constant $k$ was set to $1$, $m_1 = 4$, $m_2 = 1$, $\omega_s = 1$, $g_1 = g_2 = 1$, and $\lambda = 1$ or $0$, depending on whether the spins in the systems were coupled. The plots were generated until $T = 300$. Top left: oscillator-spin. Top right: oscillator-decoupled spins. Bottom left: oscillator-coupled spins. Bottom right: coupled oscillators.}
    \label{no_decoh_p_p}
\end{figure}

For the coupled oscillators (figure \ref{osc}), we have two $\hat h_o$ terms, along with a new coupling term, $\hat h_{oo}$:
\begin{align}
   \hat H= \hat h_{o1}+\hat h_{o2}+\hat h_{oo}.
\end{align}
The oscillator Hamiltonians are the same as in the single oscillator case. The new, third component $h_{oo}$ reads,
\begin{align}
\hat h_{oo} = \frac{\lambda}{2}(\hat x_1 \otimes I_2-\hat x_2\otimes I_2)^2.
\end{align}
We will first consider the phase plots for the standard and decohered systems. To compare the behavior of just the oscillator in all systems, we trace out the spin degrees of freedom, and for the coupled oscillator system we trace out one oscillator. The standard systems were periodic in form. Making fewer coupling parameters nonzero increased the complexity of the system, and caused more erratic phase behavior (see figures \ref{no_decoh_p_p}).

\begin{figure}[h]
    \centering
    \includegraphics[width=0.23\textwidth]{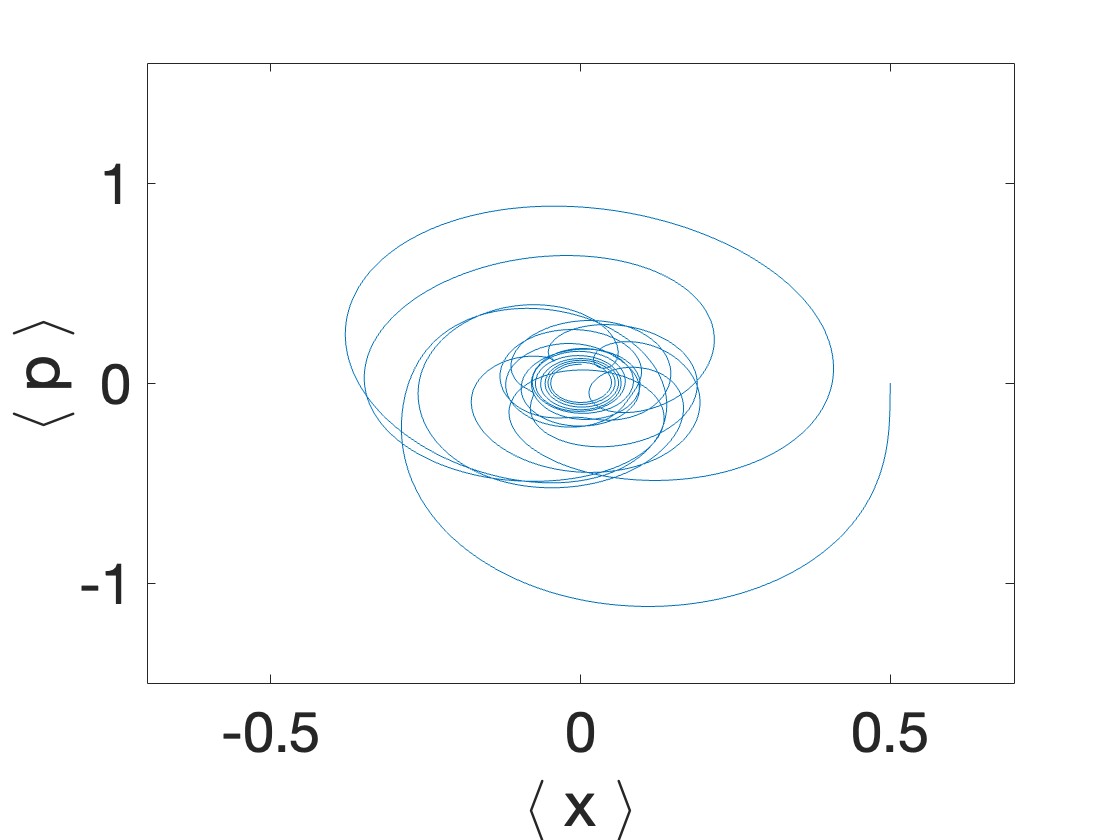} 
    \includegraphics[width=0.23\textwidth]{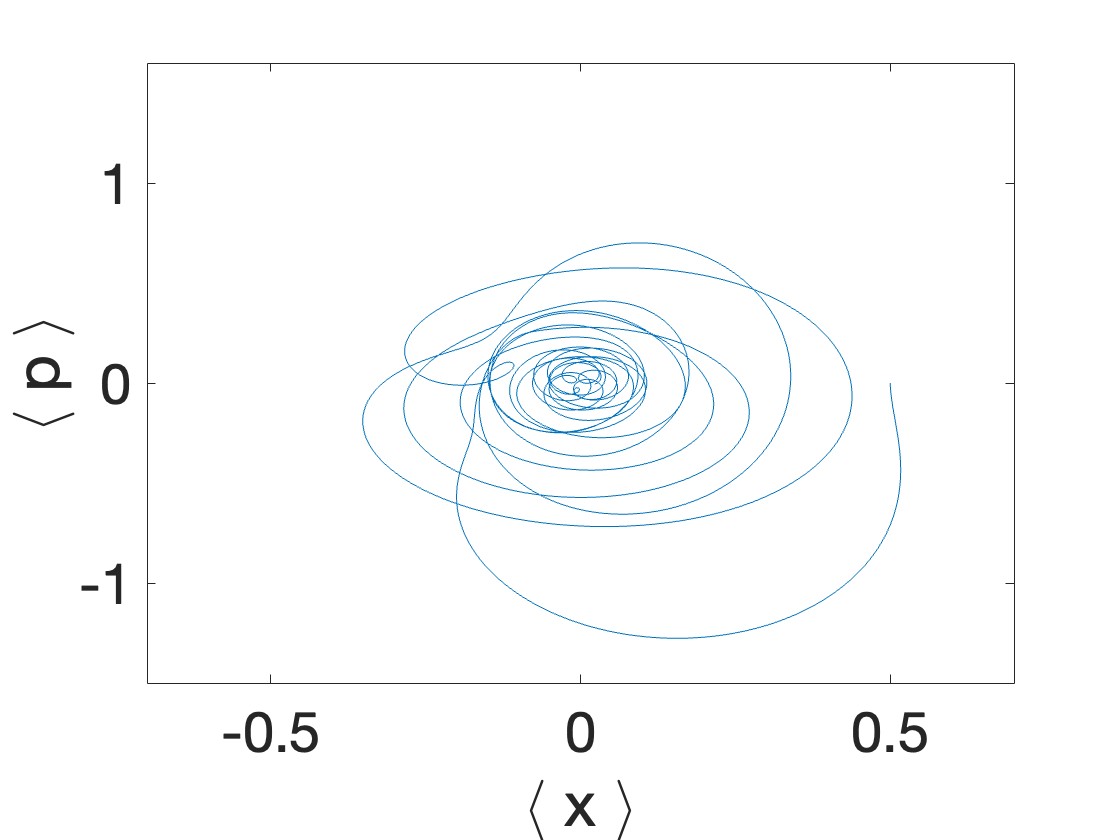} 
    \includegraphics[width=0.23\textwidth]{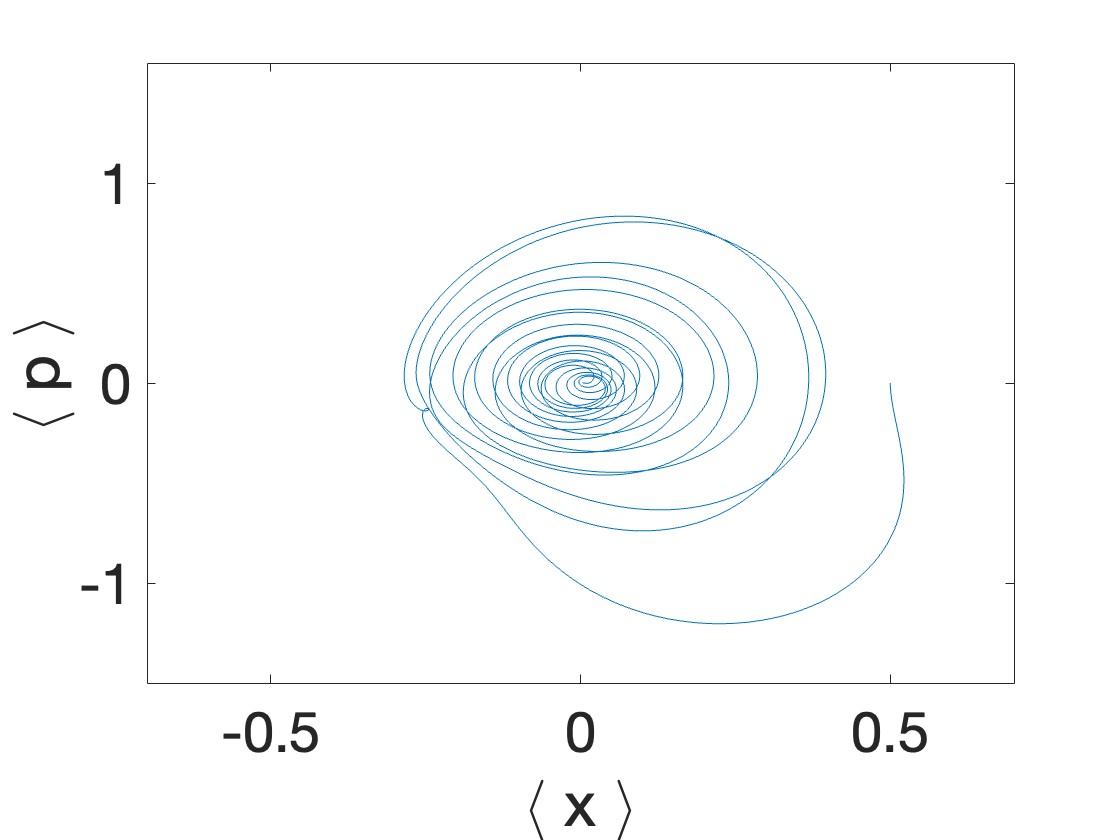} 
    \includegraphics[width=0.23\textwidth]{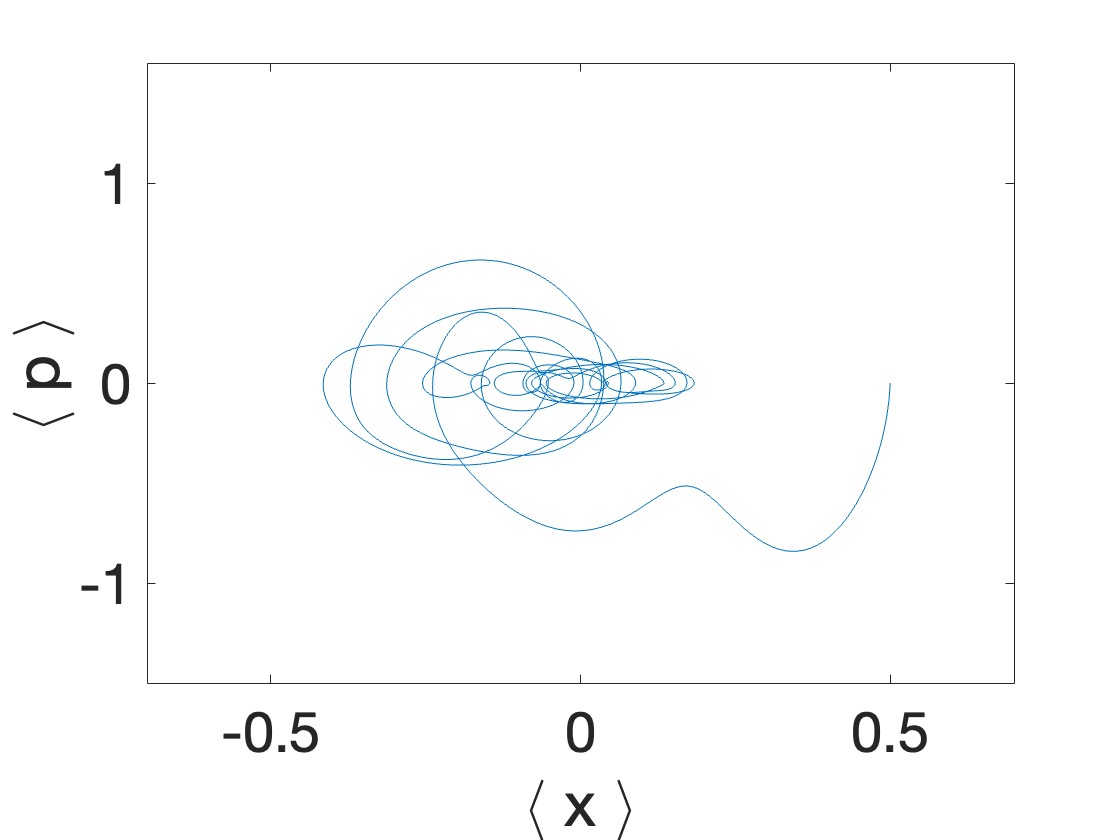} 
    \caption{Phase plot of a traced-out oscillator for the more complicated systems, with decoherence ($\tau = 0.1, x_0 = 0.5, p_0 = 0$). The spring constant $k$ was set to $1$, $m_1 = 4$, $m_2 = 1$, $\omega_s = 1$, $g_{1,2} = 1$, and $\lambda = 1$ or $0$, depending. The plots were generated until $T = 300$. Top right: oscillator-decoupled spins. Bottom left: oscillator-coupled spins Bottom right: coupled oscillators. All systems' expectation values approach zero.}
    \label{decoh_p_p}
\end{figure}

The decohered systems approached position and momentum expectation values of zero at large time scales, as in the single oscillator case, with the coupled oscillators demonstrating the most erratic behavior (see figures \ref{decoh_p_p}).
In all cases, the expectation values of the phase space variables approach zero, much like the single oscillator case. The difference arises in the complexity of the phase plots. These may be related to how energy is transferred between the oscillator and the other degrees of freedom.

\subsection{Thermodynamics, Entangled Evolutions}
\begin{figure}[h]
    \centering
    \includegraphics[width=0.49\textwidth]{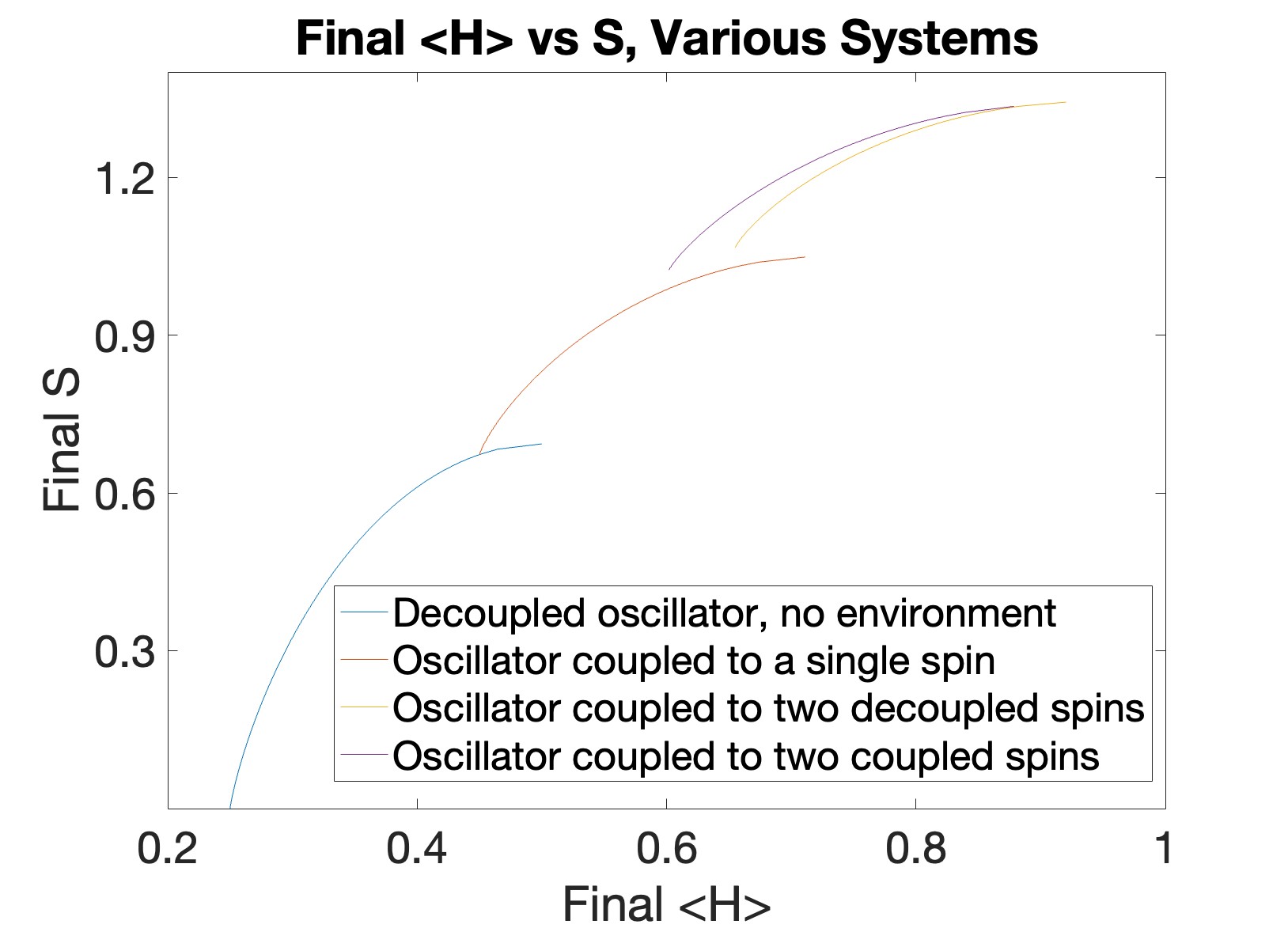} 
    \caption{Final energy expectation values for all systems after decoherence. Setting a larger initial $x_0$ value for the oscillator led to larger entropy and energy values. Note that, by (\ref{conservation}), the total energy of the system will always remain conserved. Moreover, when more energy is in the system, more entropy occurs.}
    \label{h_v_s}
\end{figure}

We will now explore links with thermodynamics, just as we did for the single oscillator case. For the oscillator coupled to two spins, the oscillator's energy is calculated via the following equation \cite{FC_Hamiltonian_Ideas}:
\begin{align}
    \braket{\hat h_o} = \braket{\hat H}- \braket{\hat h_s+\hat h_{g_1}+\hat h_{g2}+\hat h_{\lambda}}.
\end{align}
We can then compare this energy value with the final entropy of the oscillator. We sampled all allowed $x_0$ values inside of the coherent state, ranging from $x_0 = -0.5\rightarrow 0.5$. We saw again that a larger initial displacement, which corresponds to greater energy, corresponded to a larger total entropy after decoherence (see figure \ref{h_v_s}).

\begin{figure}[h]
    \centering
    \includegraphics[width=0.33\textwidth]
    {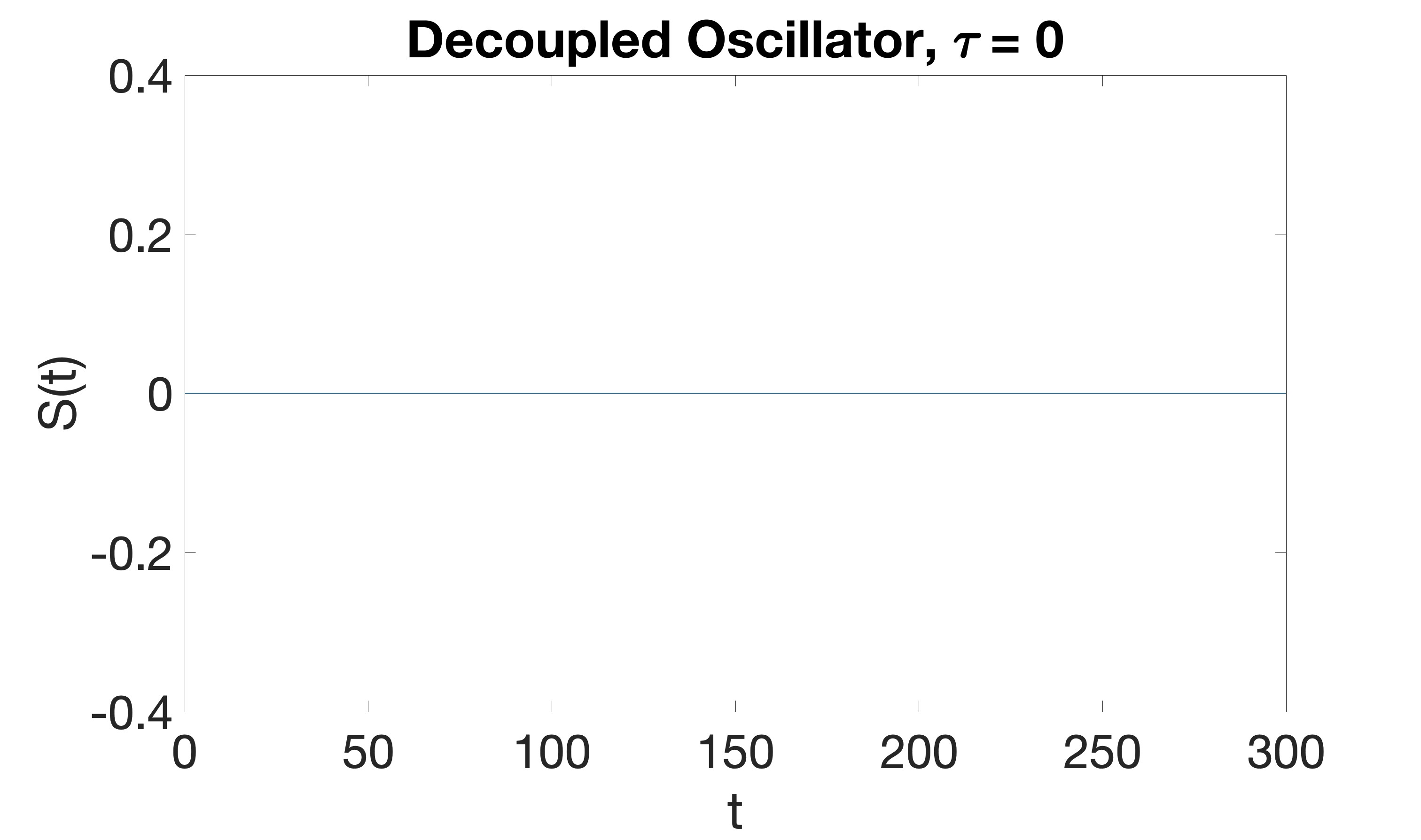}
    \includegraphics[width = 0.33\textwidth]{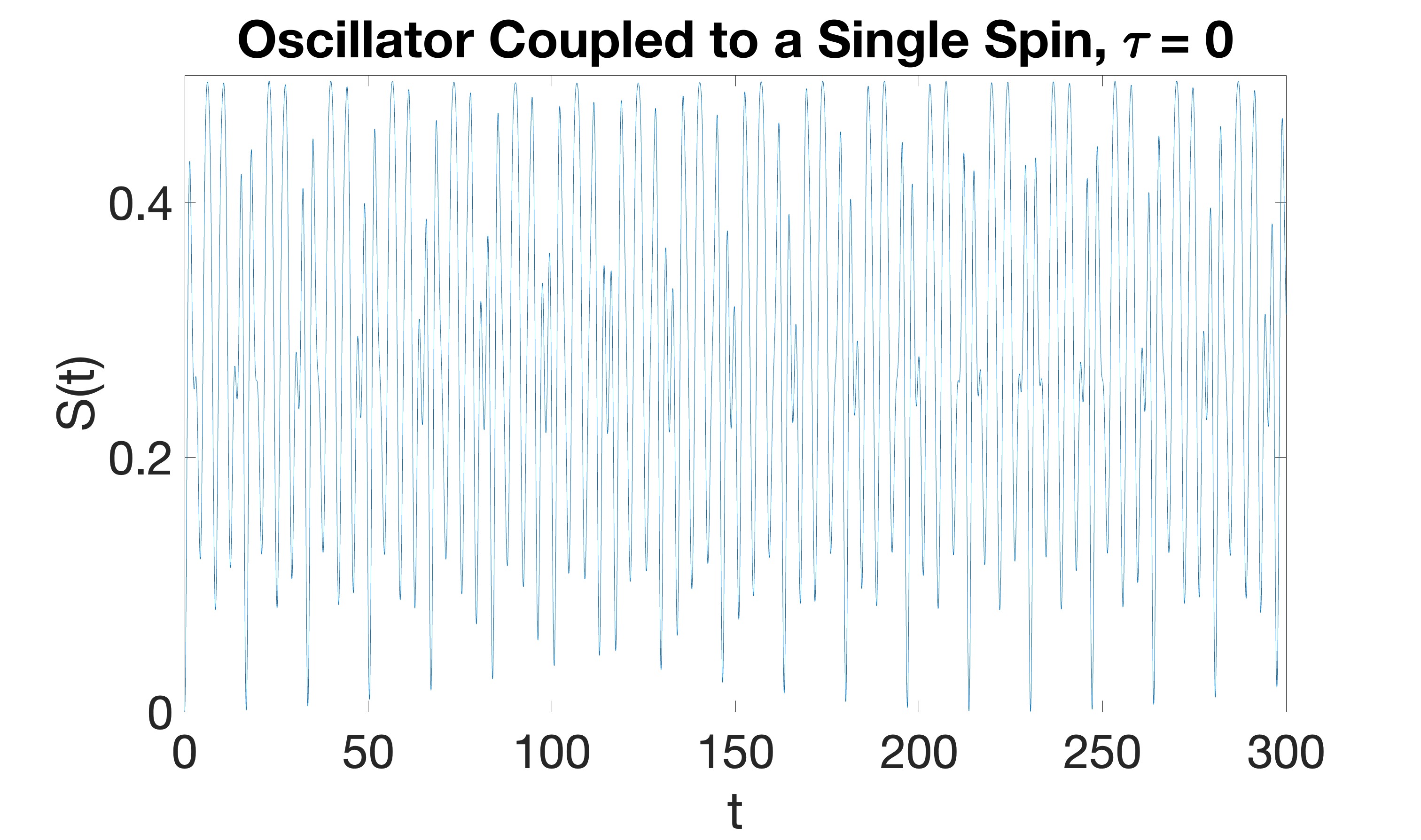}
    \includegraphics[width = 0.33\textwidth]{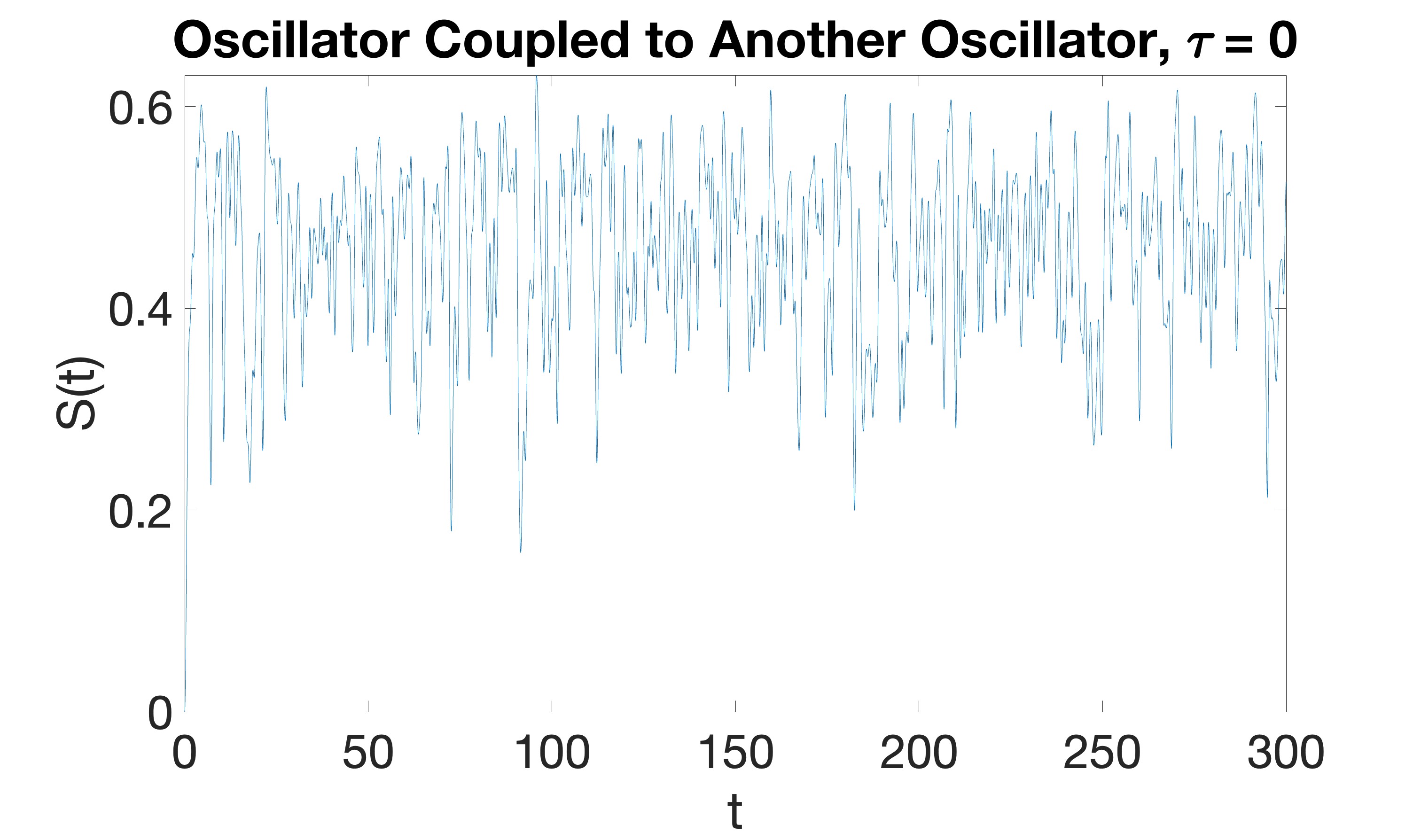}
    \includegraphics[width = 0.33\textwidth]{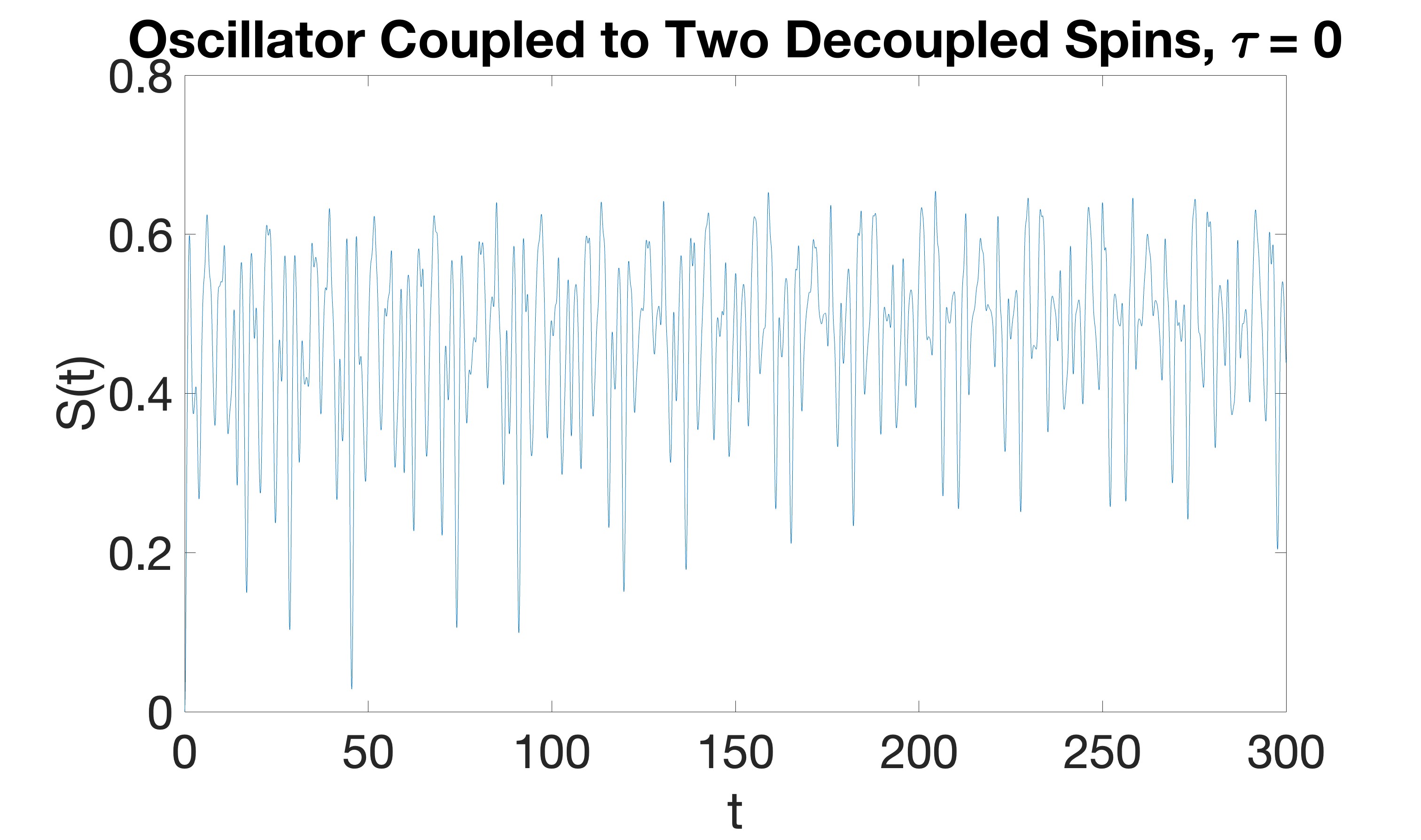}
     \includegraphics[width = 0.33\textwidth]{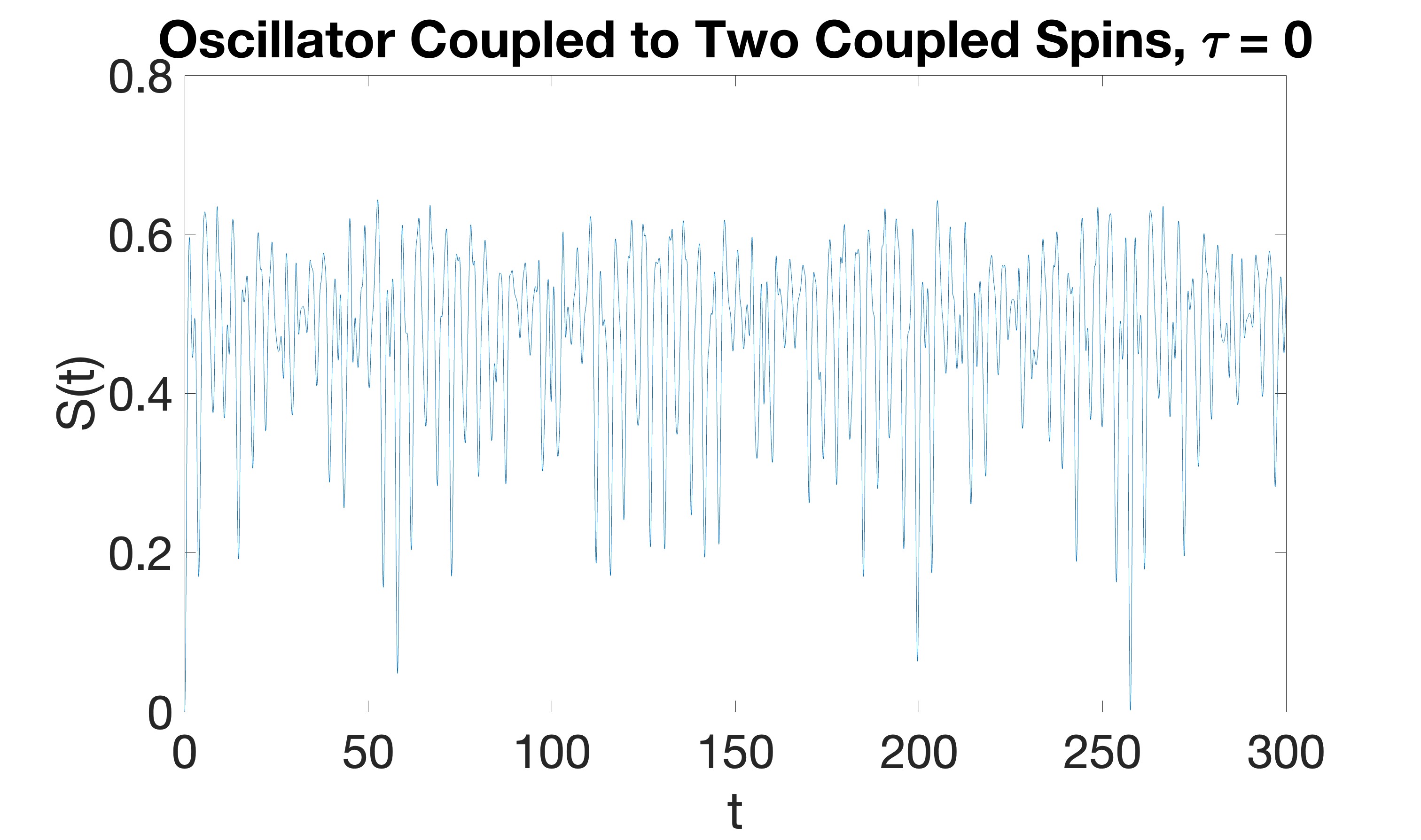}
    \caption{Entanglement entropy, without decoherence, for the five complex systems. Note that the oscillations are periodic for all systems.}
    \label{e_e}
\end{figure}

We next considered the second law---that entropy tends to increase. We first examined the behavior of the oscillators in the absence of decoherence. When the oscillator was coupled to any other element, the entropy of the system was chaotic, representing a constant exchange of information between the systems' components during evolution (see figure \ref{e_e}).

\begin{figure}[h]
    \centering
    \includegraphics[width=0.48\textwidth]{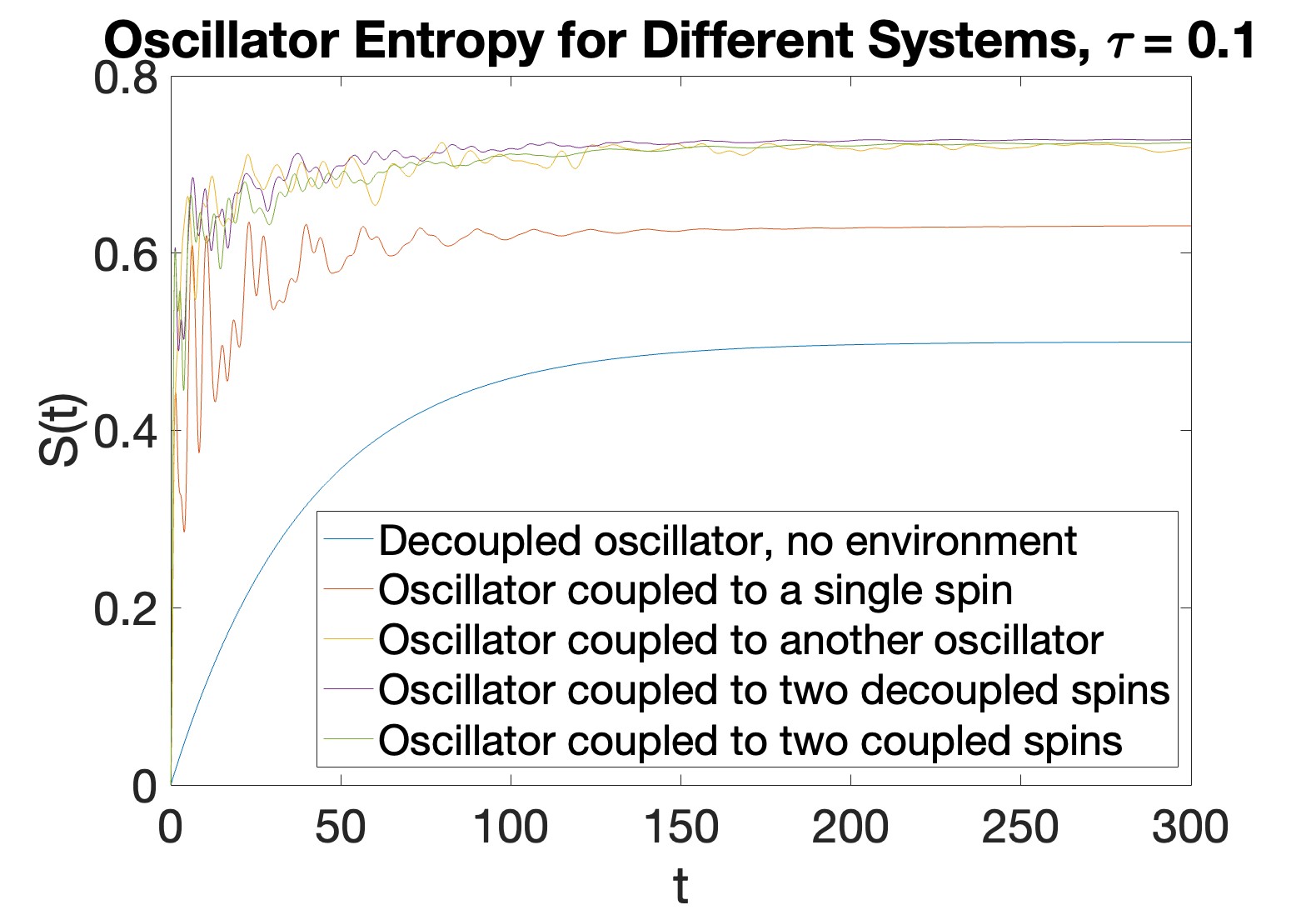} 
    \includegraphics[width=0.48\textwidth]{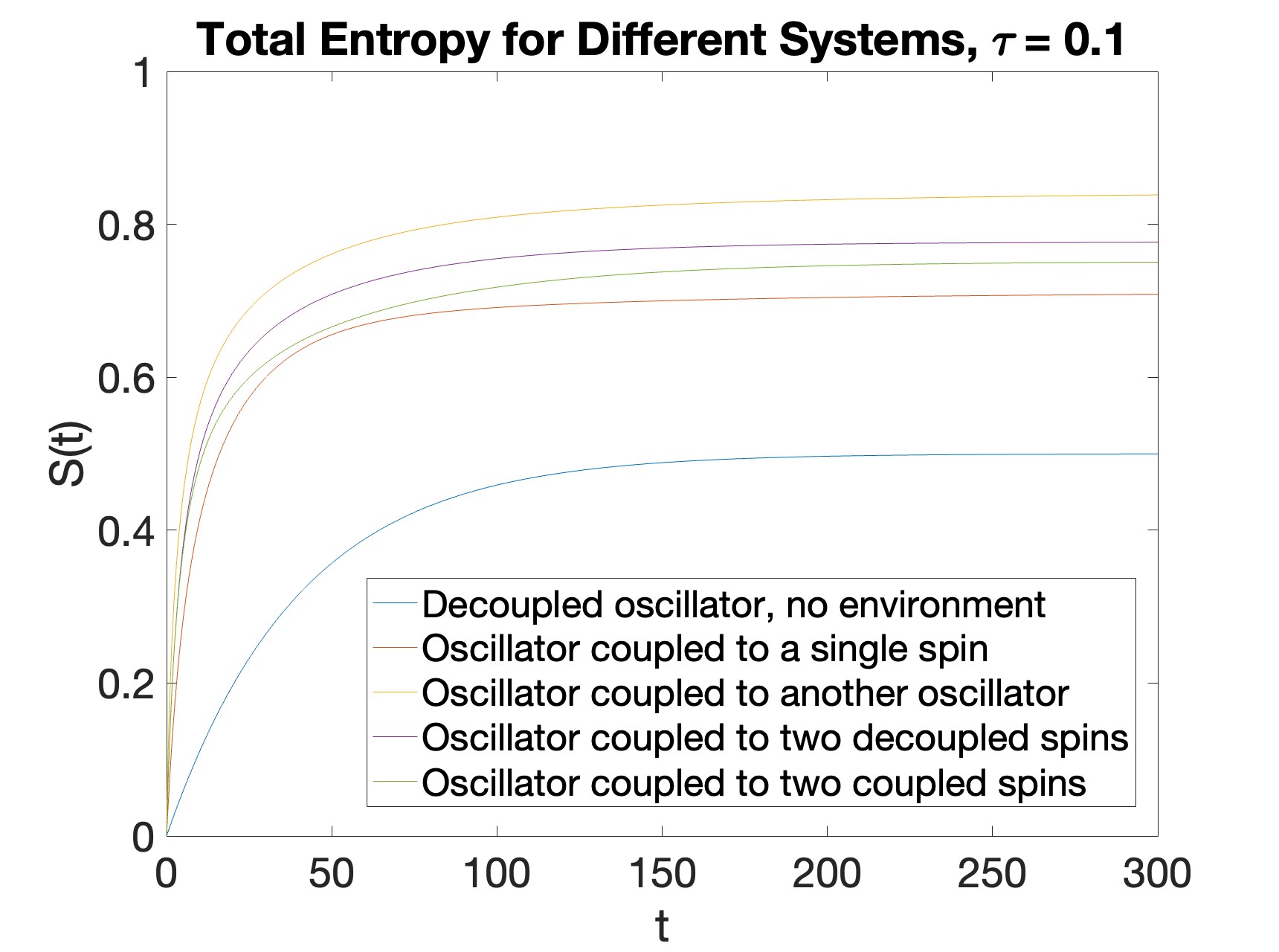}
    \caption{Entropy evolution with decoherence. Although the entanglement entropy of the oscillator does not consistently increase, that of the system as a whole does, leading to a general second law adherence.}
    \label{attractor}
\end{figure}

The entropy of systems undergoing decoherence differed. After first oscillating, similar to the standard systems, these states approach a plateau once $\tau$ dominated the state evolution. If one focuses on the entire system, however, the oscillatory behavior disappears, and the entropy constantly increases with respect to time, whereby it satisfies the second law (see figure \ref{attractor}).

Note that the interaction of spins on the oscillator is similar to environmental decoherence, and results in a higher final entropy. This can be interpreted as evidence that this system, which already naturally decoheres, will decohere further as the environment is made larger. Additionally, when the spins are uncoupled, the final decoherence of the oscillator is greater than in the fully coupled scheme, because all the energy of these spins is transferred to the oscillator, rather than partially shared between the spins.

\section{Conclusion}

This project studied an energy decoherence that occurs dynamically in closed quantum systems. The simple harmonic oscillator was considered first, and showed adherence to the first and second laws of thermodynamics. Next, more complicated systems were shown. As these systems increased in complexity, their energy and entropy values also generally increased. The increase in entropy resulted in a decoupling of components in the more complex systems, leading to a final state that resembled the decoupled oscillator case for all systems.

In the future, the results of this project will be examined under new mediums. First, by examining classical phase plots, the effect of decoherence on an oscillator with classical degrees of freedom, coupled to two spins undergoing decoherence, will be determined. Matter-gravity systems undergoing energy decoherence will also be examined. We also intend to explore whether energy decoherence occurs when one reformulates standard quantum mechanics using quantum reference frames.

{\bf Acknowledgements} This work was supported by the FDC award of Centre College; it was done in collaboration with Kraig Grauer, Irfan Javed and Mustafa Saeed.

\bibliography{Main.bib}
\end{document}